\documentclass[12pt]{article}
\let\section=\subsection     \let\subsection=\subsubsection                %%
\usepackage{graphicx}

\begin{document}
\begin{center}
   {\large \bf FINITE TEMPERATURE AND CHEMICAL}\\[2mm] 
   {\large \bf POTENTIAL IN LATTICE QCD}\\[2mm]
   {\large \bf AND ITS CRITICAL POINT}\\[5mm]
   Z. FODOR$^*$\footnote{email: fodor@poe.elte.hu} AND 
S.D. KATZ$^\ddag$\footnote{on leave from
Institute for Theoretical Physics, E\"otv\"os University,
P\'azm\'any P. 1/A, H-1117, Budapest, Hungary\\ 
\hspace*{6mm}email: katz@bodri.elte.hu} \\[5mm]
   {\small \it $^*$Institute for Theoretical Physics, E\"otv\"os University,
P\'azm\'any P. 1/A, H-1117,\\ Budapest, Hungary \\
               $^\ddag$Deutsches Elektronen-Synchrotron DESY, Notkestr. 85,
D-22607,\\ Hamburg, Germany
    \\[8mm] 
     }
\end{center}

\begin{abstract}\noindent
We propose a method to study lattice QCD at finite temperature (T) and 
chemical potential ($\mu$).
We compare the method with direct results and with the Glasgow method
by using $n_f$=4 QCD at Im($\mu$)$\neq$0.
We locate the critical endpoint
(E) of QCD on the Re($\mu$)-T plane. We use $n_f$=2+1 dynamical
staggered quarks with semi-realistic masses on $L_t$=4 lattices.
Our results are based on ${\cal{O}}(10^3-10^4)$ configurations.
\end{abstract}

QCD at finite $T$ and/or $\mu$ is of fundamental importance,
since it describes relevant features of particle physics
in the early universe, in neutron stars and in heavy ion collisions.
Extensive experimental work has been done
with heavy ion collisions at CERN and Brookhaven to explore
the $\mu$-$T$ phase boundary (cf. \cite{BS01}).  Note, that
past, present and future heavy ion
experiments with always higher and higher energies produce states
closer and closer to the $T$ axis of the $\mu$-$T$ diagram. It is
a long-standing question, whether a critical point
exists on the $\mu$-$T$ plane,
and particularly how to tell its location theoretically
\cite{crit_point}.

Let us discuss first the $\mu$=0 case.
Universal arguments \cite{PW84} and lattice simulations \cite{U97}
indicate that in a hypothetical QCD
with a strange (s) quark mass ($m_s$) as small as the up (u) and down (d)
quark masses ($m_{u,d}$)
there would be a first order finite
$T$ phase transition. The other extreme case ($n_f$=2)
with light u/d quarks but with an infinitely large $m_s$
there would be no phase transition only a
crossover. Observables change rapidly during a crossover,
but no singularities appear.
Between the two extremes there is a
critical strange mass ($m_s^c$) at which one has a second order finite
$T$ phase transition. Staggered lattice results on $L_t$=4 lattices
with two light quarks and $m_s$ around the transition $T$ ($n_f$=2+1)
indicated \cite{B90} that $m_s^c$ is about half of the physical $m_s$.
Thus, in the real world we probably have a crossover.

Returning to a non-vanishing $\mu$, one realizes that arguments
based on a variety of models (see e.g. \cite{B89,qcd_phase,crit_point})
predict a first order finite $T$ phase transition at large $\mu$.
Combining the $\mu=0$ and large $\mu$ informations an interesting
picture emerges on the $\mu$-$T$ plane. For the physical $m_s$
the first order phase transitions at large $\mu$ should be connected
with the crossover on the $\mu=0$ axis. This suggests
that the phase diagram features a critical endpoint $E$ (with
chemical potential $\mu_E$ and temperature $T_E$), at which
the line of first order phase transitions ($\mu>\mu_E$ and $T<T_E$)
ends \cite{crit_point}. At this point the phase transition is of
second order and long wavelength fluctuations appear, which
results in (see e.g. \cite{BPSS01}) consequences, similar to
critical opalescence. Passing close enough to ($\mu_E$,$T_E$)
one expects simultaneous appearance of
signatures
which exhibit nonmonotonic dependence on the
control parameters \cite{SRS99},
since one can miss the critical point on either of two sides.

The location of E is
an unambiguous, non-perturbative prediction of QCD.
No {\it ab initio}, lattice analysis based on QCD was done to locate
the endpoint.  Crude models
with $m_s=\infty$ were used (e.g. \cite{crit_point})
suggesting that $\mu_E \approx$ 700~MeV, which should be smaller
for finite $m_s$. The goal of our
exploratory work is to show how to locate the endpoint by a lattice
QCD calculation. We use full QCD with dynamical $n_f$=2+1
staggered quarks.

QCD at finite $\mu$ can be given
on the lattice \cite{HK83}; however, standard
Monte-Carlo techniques fail. At Re($\mu$)$\neq$0 
the determinant of the Euclidean Dirac operator is complex, which
spoils any importance sampling method.

Several suggestions were studied in detail to solve the problem.
We list a few of them (for a recent review see Ref. \cite{P01}).

In the large gauge coupling limit
a monomer-dimer algorithm was used \cite{KM88}.
For small gauge coupling an attractive
approach is the ``Glasgow method'' \cite{glasgow} in which the
partition function is expanded in powers of $\exp(\mu/T)$
by using an ensemble of configurations weighted by the $\mu$=0 action.
After collecting more than 20 million configurations only unphysical
results were obtained: a premature onset transition.
The reason is that the $\mu$=0 ensemble does not overlap sufficiently
with the states of interest.
Another possibility is to separate the absolute value and the
phase of the fermionic determinant and use the former to generate
configurations and the latter in observables \cite{T90}.

At imaginary $\mu$ the measure remains positive and standard Monte Carlo
techniques apply. The canonical partition function can be obtained by
a Fourier transform \cite{imag,AKW99}. In this technique the dominant
source of errors is the Fourier transform rather than the poor overlap.
One can also use the fact that the partition function away from the
transition line should be an analytic function of $\mu$, and the fit
for imaginary $\mu$ values could be analytically continued to real
values of $\mu$ \cite{L00}.  At T sufficiently above the transition,
both real and imaginary $\mu$ can be studied by
dimensionally reducing QCD \cite{HLP00}.
Hamiltonian formulation may also help studying the problem
\cite{XQLUO}.

\begin{center}
   \includegraphics[width=6.7cm,angle=0]{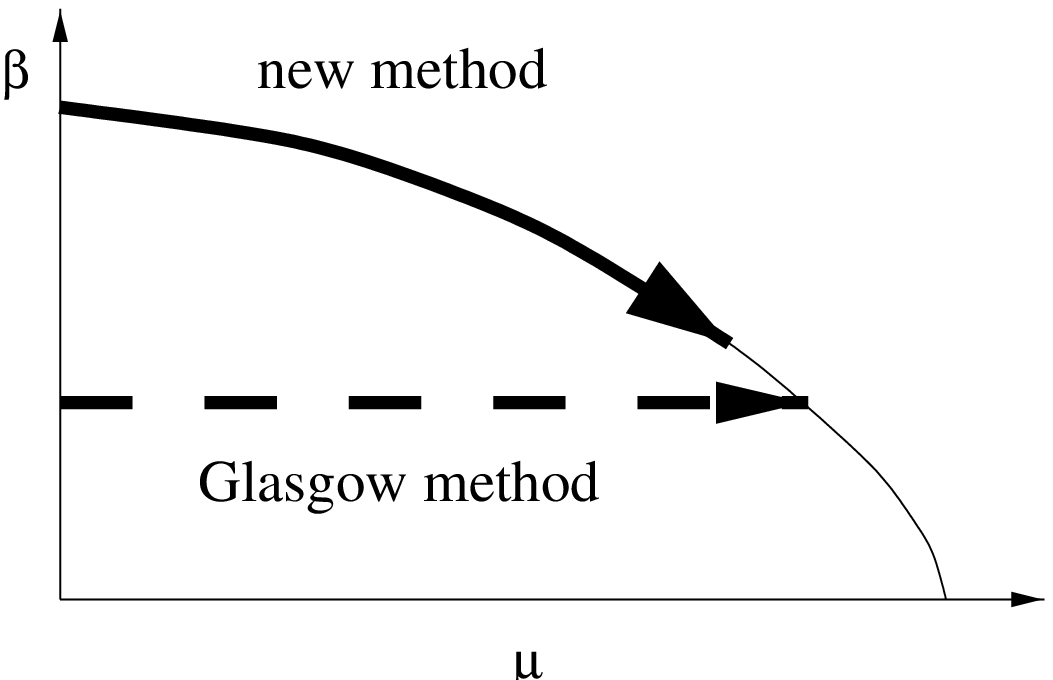}\\
   \parbox{14cm} {
\centerline
{\footnotesize 
        Fig.~1: Schematic difference between the present and the Glasgow methods.  
        }}
\end{center}

An attractive approach to alleviate the problem
is the ``Glasgow method'' (see e.g. Ref. \cite{glasgow}) in which the
partition function ($Z$) is expanded in powers of $\exp(\mu/T)$
by using an ensemble of configurations weighted by the $\mu$=0 action.
After collecting more than 20 million configurations only unphysical
results were obtained.
The reason is that the $\mu$=0 ensemble does not overlap enough
with the finite density states of interest \cite{H01}.

We propose a method
to reduce the overlap problem and determine the
phase diagram in the $\mu$-T plane (for details see \cite{FK01}).
The idea is to produce an ensemble of QCD configurations at
$\mu$=0 and at $T_c$. Then we determine
the Boltzmann weights \cite{FS89} of these configurations at $\mu\neq 0$
and at $T$ lowered to the transition temperatures at this
non-vanishing $\mu$. Since transition configurations
are reweighted to transition ones a much better
overlap can be observed than by reweighting pure had\-ronic configurations
to transition ones \cite{glasgow}. Since the original 
ensemble is collected at $\mu$=0 we do not expect to be able to
decsribe the physics of the large $\mu$ region with ie.g. exotic
colour superconductivity. Fortunately, the typical $\mu$ values
at present heavy ion accelerators are smaller than the region
we cover (for the freeze-out condition see e.g. \cite{CR99}).   

After illustrating the applicability of the method
we locate the critical point of QCD.
(Multi-dimensional reweighting
was successful for determining
the endpoint of the hot electroweak plasma \cite{ewpt}
e.g. on 4D lattices.)

Let us study a generic system of fermions $\psi$ and bosons $\phi$,
where the fermion Lagrange density is ${\bar \psi}M(\phi)\psi$.
Integrating over the Grassmann fields we get:
\begin{equation}\label{path_int}
Z(\alpha)=\int{\cal D}\phi \exp[-S_{bos}(\alpha,\phi)]\det M(\phi,\alpha),
\end{equation}
where $\alpha$ denotes a set of parameters of
the Lagrangian. In the case of staggered QCD $\alpha$
consists of $\beta$,
$m_q$ and $\mu$.
For some choice of the
parameters $\alpha$=$\alpha_0$
importance sampling can be done (e.g. for Re($\mu$)=0).
Rewriting eq. (\ref{path_int})
\begin{eqnarray}\label{reweight}
Z(\alpha)=
\int {\cal D}\phi \exp[-S_{bos}(\alpha_0,\phi)]\det M(\phi,\alpha_0)&& 
\nonumber \\
\left\{\exp[-S_{bos}(\alpha,\phi)+S_{bos}(\alpha_0,\phi)]
{\det M(\phi,\alpha)  \over \det M(\phi,\alpha_0)}\right\}.&&
\end{eqnarray}
We treat the curly bracket as an observable
(measured on each configuration)
and the rest as the measure. Changing
only one parameter of the ensemble
generated at $\alpha_0$ provides an accurate value for some observables
only for high statistics. This is ensured by
rare fluctuations as the mismatched measure occasionally sampled the
regions where the integrand is large. This is the
overlap problem. Having several parameters
the set $\alpha_0$ can be adjusted to get
a better overlap than obtained by varying only one parameter.

\begin{center}
   \includegraphics[width=7.7cm,angle=0,bb= 17 220 570 610]{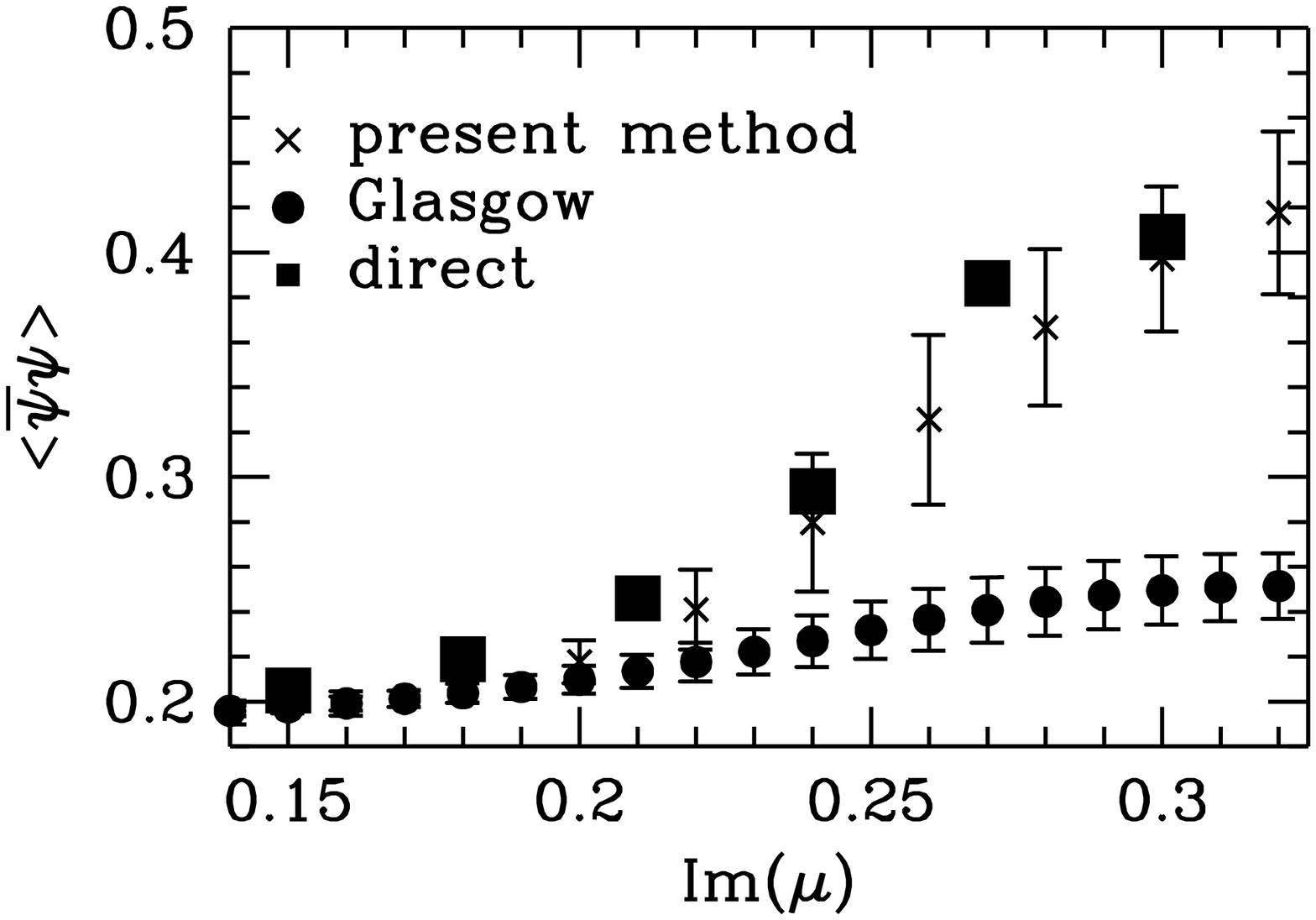}\\
   \parbox{14cm} {
%\centerline
{\footnotesize 
        Fig.~2:  ${\bar \psi}\psi$ as a
function of Im($\mu$), for direct results (squares),
our technique (crosses) and Glasgow-type reweighting (dots).
        }}
\end{center}

The basic idea of the method as applied to dynamical QCD can be
summarized as follows. We study the system at ${\rm Re}(\mu)$=0 around
its transition point. Using a Glasgow-type technique we calculate the
determinants for each configuration for a set of $\mu$, which, similarly
to the Ferrenberg-Swendsen method \cite{FS89}, can be used for
reweighting.  The average plaquette values can be used to perform an
additional reweighting in $\beta$.  Since transition configurations were
reweighted to transition ones a much better overlap can be
observed than by reweighting pure hadronic configurations to transition
ones as done by the Glasgow-type techniques. The differences between the two
methods are shown in Figure 1. Moving along the transition
line was also suggested by Ref. \cite{AKW99}.

\begin{center}
   \includegraphics[width=8.4cm,angle=0]{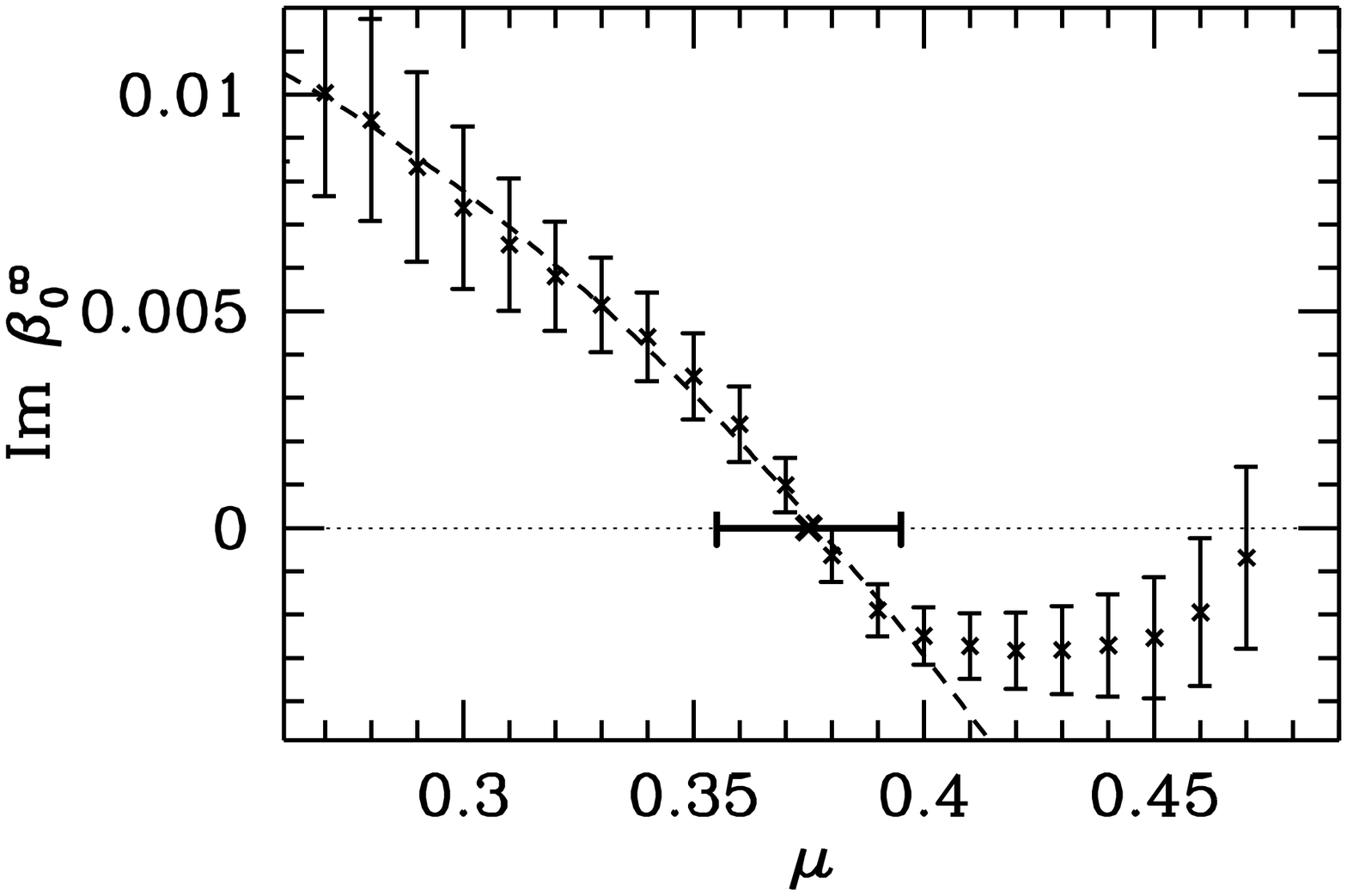}\\
   \parbox{14cm} {\centerline{\footnotesize 
        Fig.~3: Im($\beta_0^\infty$) as a function of the chemical
potential.  }}
\end{center}

We have directly tested these ideas in $n_f$=4 QCD
with $m_q$=0.05 dynamical
staggered quarks.
We first collected 1200 independent V=4$\cdot 6^3$ configurations at
Re($\mu$)=Im($\mu$)=0 and some $\beta$
values and used the Glasgow-reweighting and
also our technique to study Re($\mu$)=0, Im($\mu$)$\neq$0. At
Re($\mu$)=0, Im($\mu$)$\neq$0 direct simulations are possible.
After performing these direct simulations as well, a clear
comparison can be done. Figure 2 shows the predictions of
the three methods for the average quark condensates at $\beta$=5.085
as a function of Im($\mu$).
The predictions of our method agree with the direct results,
whereas for larger Im($\mu$) the predictions of the Glasgow
method are by several standard deviations off.
We expect that our
method can be applied at Re($\mu$)$\neq$0.

In QCD with $n_f$ staggered quarks
one should change the determinants to their $n_f$/4 power in our two
equations. Importance sampling works also in this case  at some $\beta$ and
at Re($\mu$)=0. Since $\det M$ is complex
an additional problem arises, one should
choose among the possible Riemann-sheets of the fractional power
in eq. (\ref{reweight}). This can be done by using \cite{FK01}
the fact that at $\mu$=$\mu_w$ the ratio of the determinants is 1 and
it should be a continuous function of $\mu$.

In the
following we keep $\mu$ real and look for the zeros of $Z$
for complex $\beta$.  At a first order phase transition the free
energy $\propto \log Z(\beta)$ is non-analytic.
A phase transition appears only in the V$\rightarrow \infty$ limit,
but not in a finite $V$. Nevertheless, $Z$
has zeros at finite V, generating the non-analyticity of the
free energy, the Lee-Yang zeros \cite{LY52}.
These are at complex parameters (e.g. $\beta$). For a
system with first order transition these zeros
approach the real axis as V$\rightarrow \infty$ 
by a $1/V$ scaling.
This V$\rightarrow \infty$ limit generates the non-analyticity of
the free energy. For a system with crossover
$Z$ is analytic, and the zeros do
not approach the real axis as V$\rightarrow \infty$.

\begin{center}
   \includegraphics[width=7.7cm,angle=0,bb= 17 210 570 610]{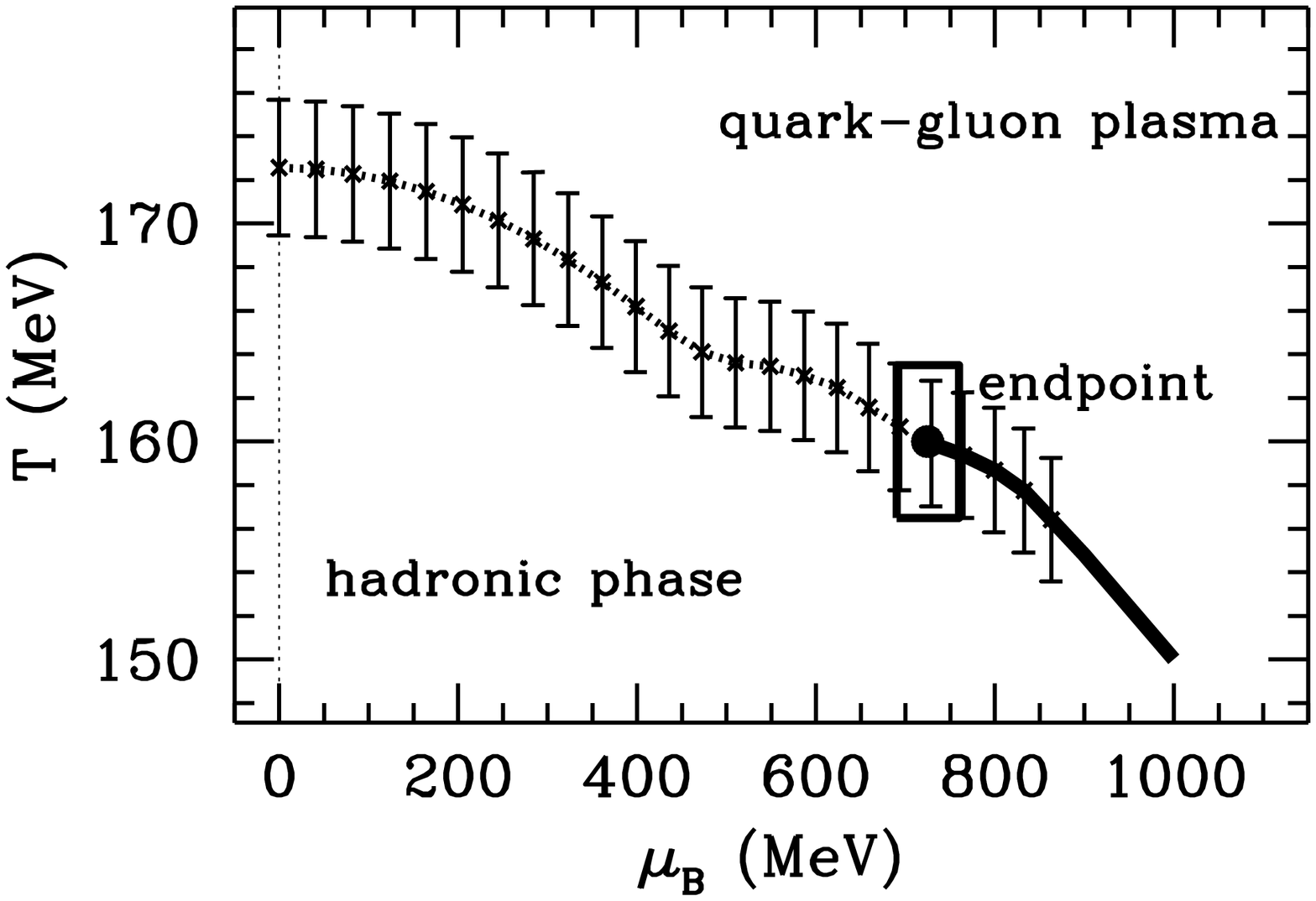}\\
   \parbox{14cm} {
%\centerline
{\footnotesize 
        Fig.~4: The T-$\mu$ diagram. Direct results are given with errorbars.
Dotted line shows the crossover, solid line the first order
transition. The box gives the uncertainties of the endpoint.
        }}
\end{center}

At T$\neq$0 we used $L_t$=4, $L_s$=4,6,8 lattices. T=0 runs were done on
$10^3\cdot$ 16 lattices. $m_{u,d}$=0.025 and $m_s$=0.2 were
our bare quark masses.
At  $T\neq 0$ we determined the complex valued Lee-Yang zeros,
$\beta_0$, for different V-s as a function of $\mu$. Their
V$\rightarrow \infty$ limit was given by a $\beta_0(V)=\beta_0^\infty+\zeta/V$
extrapolation. We used 14000, 3600 and 840 configurations on
$L_s$=4,6 and $8$ lattices, respectively.
Im($\beta_0^\infty$) is shown on Figure 3 as a function of $\mu$.  For small
$\mu$-s the extrapolated Im($\beta_0^\infty$) is inconsistent with
a vanishing value, and predicts a crossover.
Increasing $\mu$ the value of Im($\beta_0^\infty$) decreases,
thus the transition becomes consistent with a first order phase
transition (overshooting is a finite V effect).
Our primary result is $\mu_{end}=0.375(20)$.

To set the physical scale we used a
weighted average of $R_0$,  $m_\rho$  and
$\sqrt{\sigma}$.
Note, that (including systematics due to
finite V) we have
$(R_0\cdot m_\pi)=0.73(6)$, which is at least twice, $m_{u,d}$ is
at least four times
as large as the physical values.

Figure 4 shows the phase diagram in
physical units, thus
$T$ as a function of $\mu_B$, the baryonic chemical potential
(which is three times larger then the quark chemical potential).
The endpoint
is at $T_E=160 \pm 3.5$~MeV, $\mu_E=725 \pm 35$~MeV.
At $\mu_B$=0 we obtained $T_c=172 \pm 3$~MeV.

We proposed a method --an overlap improving multi-parameter reweighting
technique-- to numerically study non-zero $\mu$ and determine the
phase diagram in the $T$-$\mu$ plane.
Our method is applicable to any number of Wilson or staggered quarks.
As a direct test we showed that for Im($\mu$)$\neq$0 the predictions
of our method are
in complete agreement with the direct simulations, whereas the Glasgow
method suffers from the well-known overlap problem.
We studied the $\mu$-$T$ phase diagram of QCD with
dynamical $n_f$=2+1 quarks.
Using our method we obtained
$T_E$=160$\pm$3.5~MeV and $\mu_E$=725$\pm$35~MeV for the endpoint.
Though $\mu_E$ is too
large to be studied at RHIC or LHC, the endpoint would
probably move closer to the $\mu$=0 axis
when the quark masses get reduced.
At $\mu$=0 we obtained $T_c$=172$\pm$3~MeV.
More work is needed to get
the final values by extrapolating
in the R-algorithm and to the thermodynamic, chiral and continuum limits.
The details of the presented results can be found in \cite{FK01}.

This work was partially supported by 
grants 
 OTKA-\-T34980/\-T29803/\-M37071/\-OM-MU-708/\-IKTA111/\-NIIF
and in part based
on the MILC collaboration's  lattice code:
http://physics.indiana.edu/\~{ }sg/milc.html.

\end{document}